\documentclass[aps,prb,superscriptaddress,twocolumn,floatfix,notitlepage,a4paper]{revtex4-1}

\usepackage{graphicx}
\usepackage{amsmath,amssymb,amsfonts}
\usepackage{bm}
\usepackage{color}
\usepackage[dvipsnames]{xcolor}
\usepackage[percent]{overpic}
\usepackage{paralist}
\usepackage[breaklinks=true,colorlinks,citecolor=blue,linkcolor=blue,urlcolor=blue]{hyperref}
\usepackage{comment}
\DeclareMathOperator{\sign}{sign}

\begin{document}

\title{Magnetic hallmarks of viscous electron flow in graphene}

\author{Karina A. Guerrero-Becerra}
\affiliation{Istituto Italiano di Tecnologia, Via Morego 30, 16163 Genova, Italy}

\author{Francesco M. D. Pellegrino}
\affiliation{Dipartimento di Fisica e Astronomia, Universit\`{a} di Catania, Via S. Sofia, 64, I-95123 Catania, Italy}
\affiliation{INFN, Sez. Catania, I-95123 Catania, Italy}

\author{Marco Polini}
\affiliation{Istituto Italiano di Tecnologia, Via Morego 30, 16163 Genova, Italy}
\affiliation{School of Physics \& Astronomy, University of Manchester, Oxford Road, Manchester M13 9PL, United Kingdom}

\begin{abstract}
We propose a protocol to identify spatial hallmarks of viscous electron flow in graphene and other two-dimensional viscous electron fluids. We predict that the profile of the magnetic field generated by hydrodynamic electron currents flowing in confined geometries displays unambiguous features linked to whirlpools and backflow near current injectors. 
We also show that the same profile sheds
light on the nature of the boundary conditions describing friction exerted on the electron fluid by the edges of the sample. 
Our predictions are within reach of vector magnetometry based on nitrogen-vacancy centers embedded in a diamond slab mounted onto a graphene layer.
\end{abstract}
\maketitle
\textit{Introduction.}---Electrical transport~\cite{bandurin_science_2016,moll_science_2016,kumar_natphys_2017,berdyugin_arxiv_2018,bandurin_arxiv_2018}, thermal transport~\cite{crossno_science_2016}, and scanning gate spectroscopy~\cite{braem_arxiv_2018} measurements have recently been used to identify signatures of viscous electron flow in high-quality graphene, palladium cobaltate, and GaAs. (For a recent review see e.g.~Ref.~\onlinecite{lucas_jphyscondens_2018}.) In this regime of transport dominated by electron-electron interactions, viscosity determines electron whirlpools in the steady-state current pattern, which have been theoretically studied with great detail and are expected to emerge in confined geometries~\cite{torre_prb_2015,pellegrino_prb_2016,pellegrino_prb_2017,levitov_naturephys_2016}.
So far, a direct experimental observation of electron whirlpools and associated backflow near current injectors is still lacking. 

A promising route to achieve real space imaging of spatial patterns of current flow in two-dimensional (2D) materials is to employ vector magnetometry based on nitrogen-vacancy (NV) centers in diamond~\cite{casola_naturereview_2018}, which combines the benefits of high spatial resolution and competitive magnetic field resolution. NV vector magnetometry optically detects the field-dependent magnetic resonances of an ensemble of NV centers, from which, relying on schemes based on an external magnetic field~\cite{maertz_applphyslett_2010,tetienne_science_2017}, optical polarization~\cite{munzhuber_arxiv_2017} or Fourier optical decomposition~\cite{backlund_pra_2017}, the Cartesian components of the local magnetic field are determined.
The capability of this noninvasive imaging technique to access the details of 2D spatial flow patterns has been recently demonstrated in graphene in the diffusive regime~\cite{tetienne_science_2017}. NV vector magnetometry operates over a wide range of temperatures~\cite{acosta_prl_2010}, including room temperature~\cite{taylor_naturephys_2008}, and its spatial resolution is comparable with the viscosity diffusion length in graphene~\cite{torre_prb_2015,pellegrino_prb_2016,pellegrino_prb_2017,levitov_naturephys_2016,bandurin_science_2016,kumar_natphys_2017,berdyugin_arxiv_2018,bandurin_arxiv_2018}. Recently, the electronic spin of a {\it single} NV center attached to a scanning tip and operated under ambient conditions was used to image and detect microwave fields in a micron-scale stripline~\cite{appel_njp_2015} and to image charge flow in carbon nanotubes and Pt nanowires~\cite{chang_nanolett_2017}.

\begin{figure}
\begin{overpic}[width=\linewidth]{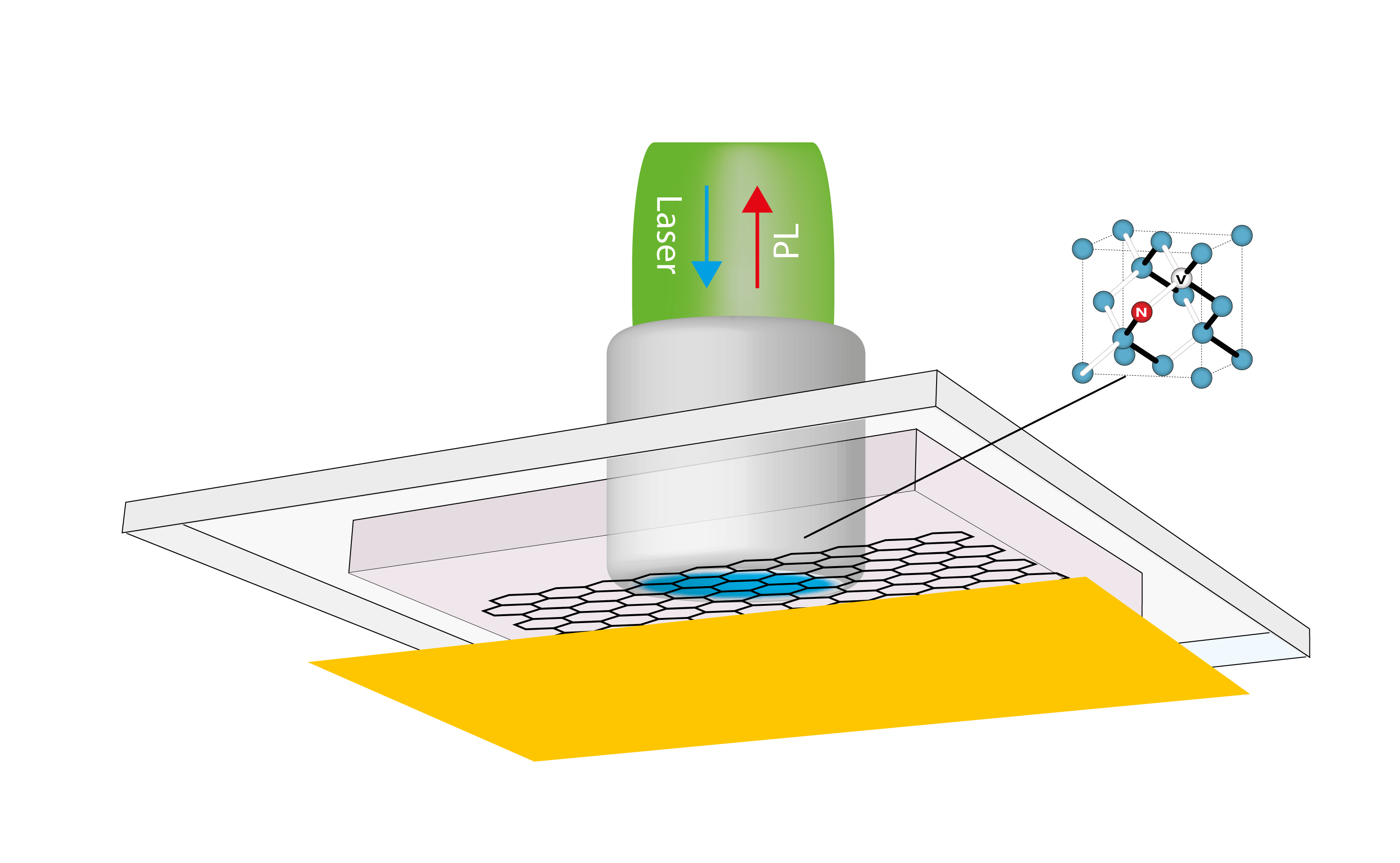}\end{overpic}
\caption{(Color online) Schematic representation of an NV magnetometry experiment aimed at detecting electron whirlpools in graphene. 
A graphene rectangular sample (hexagonal lattice) is mounted onto a diamond slab (pale-pink slab) hosting an array of NV centers (zoom). 
A back gate (yellow) controls the equilibrium electron density in graphene. 
(Graphene encapsulation in e.g. hexagonal Boron Nitride to ensure high electronic quality is not shown.) 
An optical setup (structure sketched above the diamond slab) prepares the quantum state of the NV centers through a laser beam and accesses its photoluminescence (PL) as a 
function of the frequency of a microwave excitation (not shown). 
This setup enables measurements of the magnetic field ${\bm B}({\bm x}, z)$ generated by the current density flowing in graphene 
and a spatial map of the 2D current density ${\bm J}({\bm x})$ can be reconstructed from the Cartesian components of ${\bm B}$. \label{fig:setup} }
\end{figure}

In this Rapid Communication we propose to apply NV vector magnetometry to detect viscous spatial flow patterns in graphene. We calculate the magnetic field generated by hydrodynamic currents flowing in a graphene sample of rectangular shape, placed below an array of NV centers and above a metallic back gate. A cartoon of the geometry is shown in Fig.~\ref{fig:setup}. We show that this field carries unambiguous signatures of electron whirlpools. 

\textit{Transport equations in viscous 2D electron systems.---}In the linear response regime and in a steady-state, viscous electron transport in a 2D electron fluid is described~\cite{torre_prb_2015,pellegrino_prb_2016,pellegrino_prb_2017,levitov_naturephys_2016} by the continuity
\begin{equation} \label{eq:continuity}
\bm{\nabla} \cdot \bm{J}(\bm{x}) = 0 \, ,
\end{equation}
and Navier-Stokes
\begin{equation} \label{eq:ns}
D_{\nu}^{2} \nabla^2 \bm{J}(\bm{x}) -\sigma_{0} \nabla \phi(\bm{x})= \bm{J}(\bm{x})
\end{equation}
equations. Here, $\bm{x}=(x,y)$ describes the position in the plane where electrons roam, $\bm{J}(\bm{x})$ the current density, $\phi(\bm{x})$ the 2D electrostatic potential, and the characteristic viscosity diffusion length $D_{\nu}= \sqrt{\nu \tau}$ has been introduced in Ref.~\onlinecite{torre_prb_2015}, $\nu$ being the kinematic shear viscosity and $\tau$ a phenomenological transport time describing momentum-non-conserving collisions. In Eq.~(\ref{eq:ns}), $\sigma_{0} \equiv e^2 \overline{n} \tau/m$ is a Drude-like conductivity, where
$\overline{n}$ denotes the equilibrium electron density, which can be controlled by a metallic gate, and $m= \hbar k_{\rm F}/v_{\rm F}$ is the electron effective mass in graphene, 
with $k_{\rm F}= \sqrt{\pi \overline{n}}$ and $v_{\rm F}\simeq 10^{6}~{\rm m}/{\rm s}$ the Fermi wave number and the Fermi velocity, respectively. 
Eqs.~(\ref{eq:continuity}) and (\ref{eq:ns}) can be solved for $\bm{J}(\bm{x})$ and $\phi(\bm{x})$ 
by introducing suitable boundary conditions~\cite{torre_prb_2015,pellegrino_prb_2016,pellegrino_prb_2017,levitov_naturephys_2016}.
These solutions will be used below to study signatures of viscous electron flow, as carried by the magnetic field generated by ${\bm J}({\bm x})$.
\begin{figure}[t!]
\begin{overpic}[width= \linewidth]{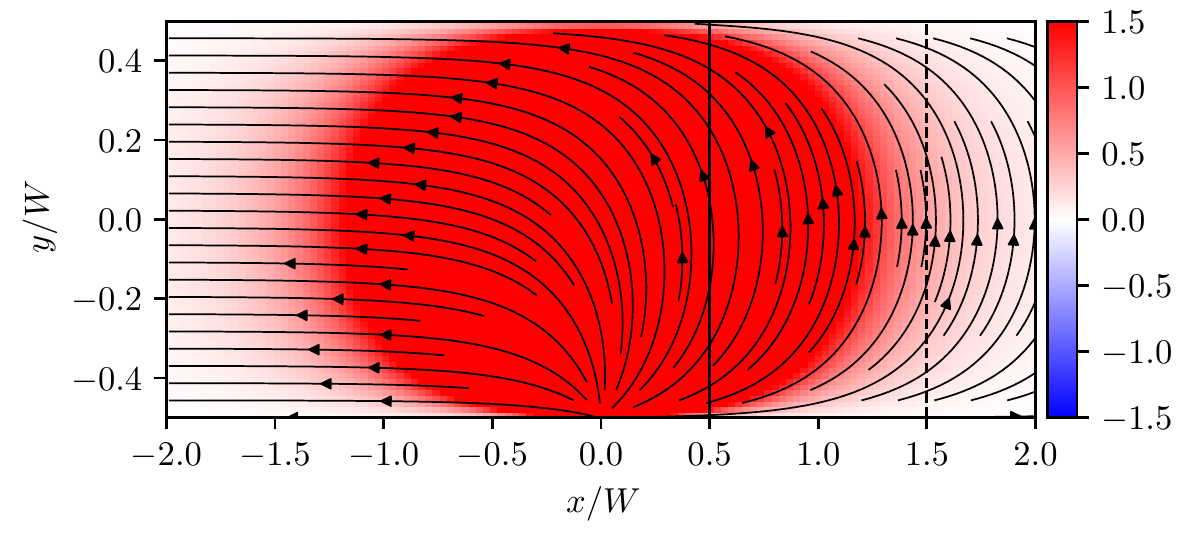}\put(1.85,45){(a)}\end{overpic}
\begin{overpic}[width= \linewidth]{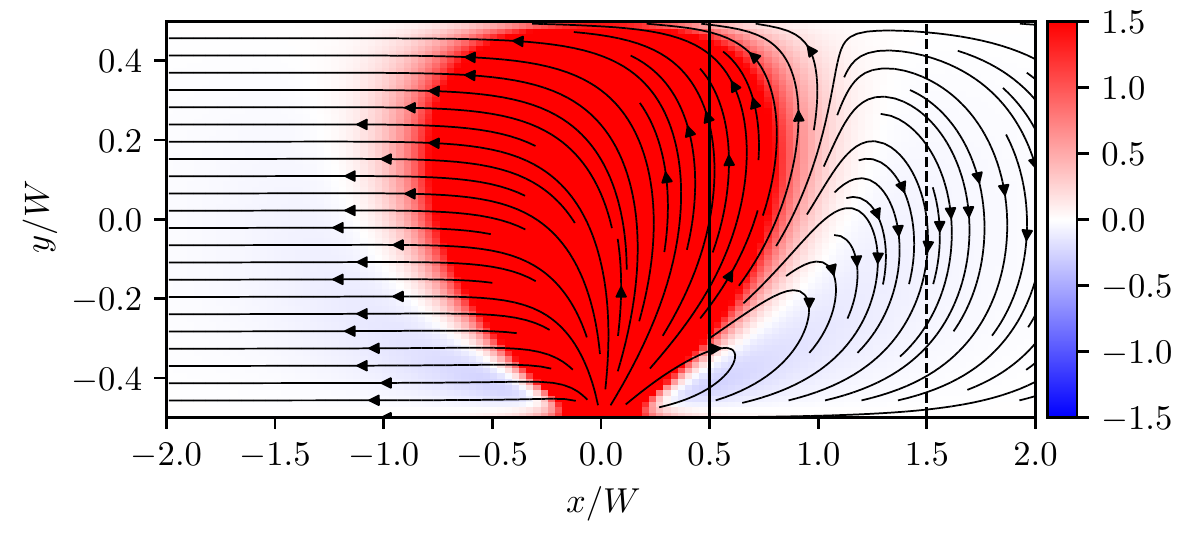}\put(1.85,45){(b)}\end{overpic}
\caption{\label{fig:vicinity-geometry-far-col-bx} 
(Color online) Spatial map of the ${\hat {\bm x}}$ component, $B_{x}({\bm x}, z=d')$, of the magnetic field (indicated by the color map and in ${\rm \mu T}$) 
generated by $J_{y}({\bm x})$. The current density is represented by the vector field. Panel (a) Ohmic case, $D_{\nu}= 0$. Panel (b) Viscous case, $D_{\nu}= W/4$.  
The injector is located at position $(0,-W/2)$. The collector (not shown) is set on the lower edge of the strip, at position $( -20 \, W,-W/2)$. 
An electron whirlpool is clearly seen to the right of the injector in the viscous case. 
The vertical solid and dashed lines denote the horizontal positions where one-dimensional cuts of $B_{x}({\bm x},z=d')$ have been taken---see Fig.~\ref{fig:vicinity-geometry-far-col-bx-cut}.} 
\end{figure}

\textit{Magnetic field generated by 2D current profiles.---}A 2D current density, $\bm{J}(\bm{x})$, confined at $z=0$, above a metallic gate placed at $z=-d$, generates a magnetic field at $z>0$, 
$\bm{B}(\bm{x},z)= [B_{x}(\bm{x}, z), B_{y}(\bm{x}, z), B_{z}(\bm{x}, z)]$, with components~\cite{poisson-eq-solution}
\begin{equation}  \label{eq:bx}
B_{x}(\bm{x}, z) = \frac{\mu_{0}}{2}\int  d^2\bm{x'} \mathcal{K}_{xy}(\bm{x}-\bm{x'},z)J_{y}(\bm{x'}) ~, 
\end{equation}
\begin{equation} \label{eq:by}
B_{y}(\bm{x}, z) = -\frac{\mu_{0}}{2}\int d^2\bm{x'}\mathcal{K}_{xy}(\bm{x}-\bm{x'},z)J_{x}(\bm{x'}) ~,
\end{equation}
and 
\begin{equation} \label{eq:bz}
B_{z}(\bm{x}, z) = \frac{\mu_{0}}{2}\int d^2\bm{x'} \mathcal{K}_{z}(\bm{x}-\bm{x'},z)[\nabla \times \bm{J}(\bm{x'})]_{z} ~.
\end{equation}
The kernels appearing in the above convolutions read as following: 
\begin{align} \label{eq:kernel-xy}
\mathcal{K}_{xy}(\bm{x}-\bm{x'},z) &\equiv \dfrac{z}{2 \pi [|\bm{x-x'}|^2+ z^2]^{3/2}}
\end{align} 
and
\begin{align}  \label{eq:kernel-z}
\mathcal{K}_{z}(\bm{x}-\bm{x'},z) &\equiv \dfrac{1}{2 \pi [|\bm{x-x'}|^2+ z^2]^{1/2}}~, 
\end{align} 
and $\mu_{0}$ denotes the free-space permeability.
The nonlocal relations~\cite{nonlocal-rel} in Eqs.~(\ref{eq:bx}),~(\ref{eq:by}), and~(\ref{eq:bz}) are obtained by solving the Poisson equation for the vector potential, accounting for appropriate boundary conditions~\cite{poisson-eq-solution}.  Eqs.~(\ref{eq:bx}),~(\ref{eq:by}), and~(\ref{eq:bz}) describe a one-to-one correspondence between a generic 2D current density and the generated magnetic field.

\begin{figure}[t!]
\begin{overpic}[width= 0.49\linewidth]{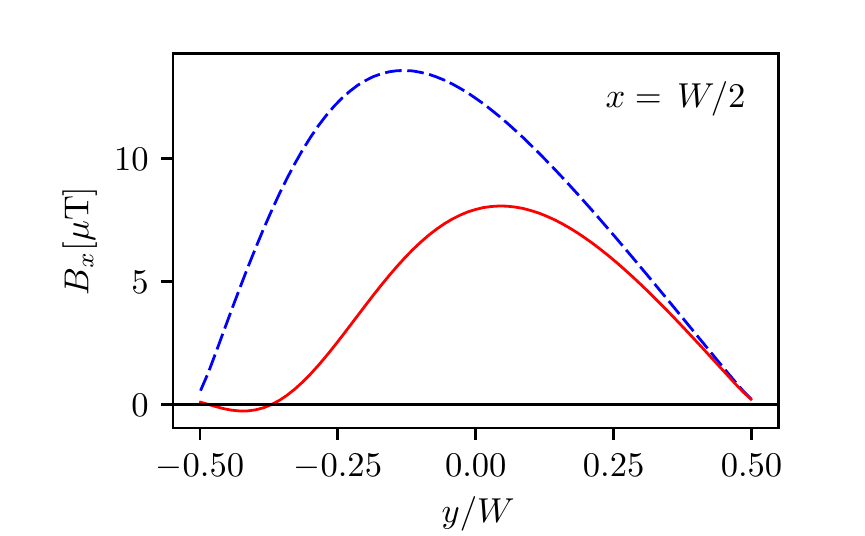}\put(1.6,60){(a)}\end{overpic}
\begin{overpic}[width= 0.49\linewidth]{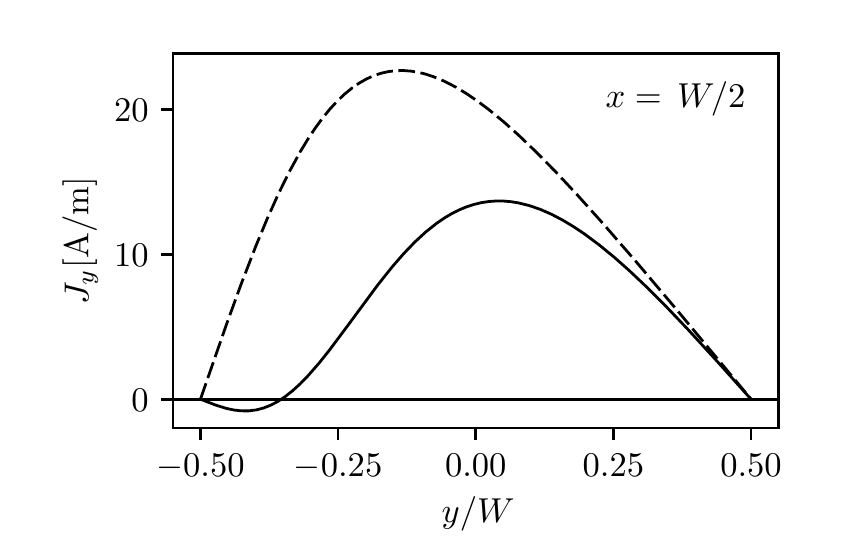}\put(1.8,60){(b)}\end{overpic}
\begin{overpic}[width= 0.49\linewidth]{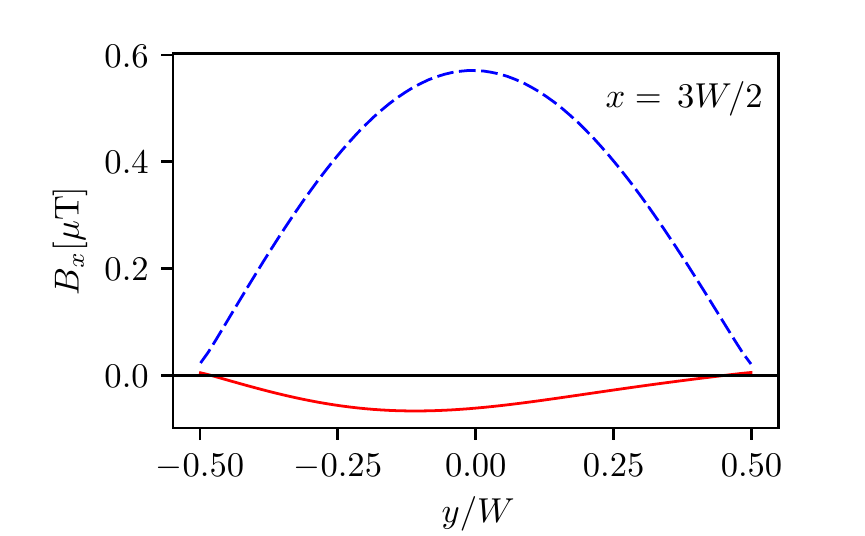}\put(1.6,60){(c)}\end{overpic}
\begin{overpic}[width= 0.49\linewidth]{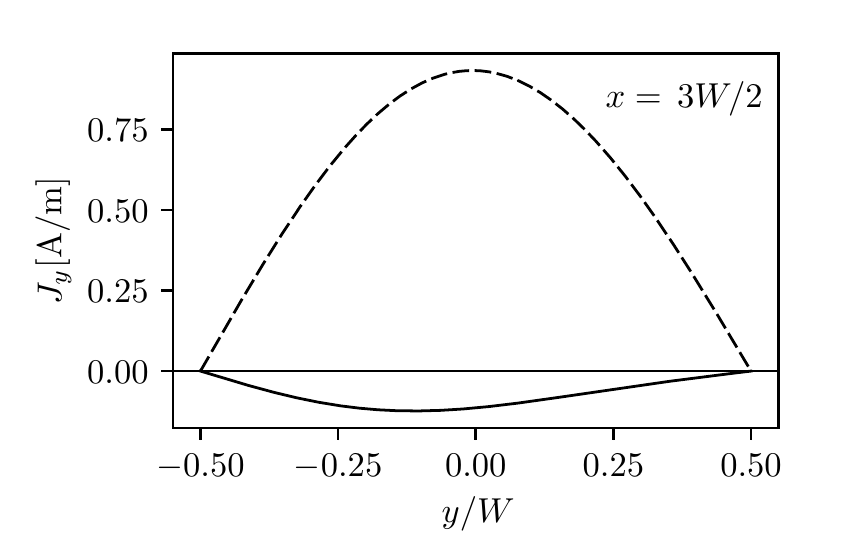}\put(1.8,60){(d)}\end{overpic}
\caption{\label{fig:vicinity-geometry-far-col-bx-cut}
(Color online) One-dimensional cuts of the 2D spatial maps reported in Fig.~\ref{fig:vicinity-geometry-far-col-bx}.
Panels (a) and (c) [(b) and (d)] illustrate $B_{x}({\bm x}, z=d')$ (in units of ${\rm \mu T}$) [$J_{y}({\bm x})$ (in units of ${\rm A}/{\rm m}$)] as a function of $y/W$, evaluated at $x =  W/2$ and $x =  3W/2$, respectively. These horizontal positions have been marked by a solid and a dashed line in Fig.~\ref{fig:vicinity-geometry-far-col-bx}, respectively.   
Solid lines in this plot correspond to the viscous case ($D_{\nu}= W/4$), while dashed lines correspond to the Ohmic case ($D_{\nu}=0$).} 
\end{figure}
\textit{Magnetic hallmarks of viscous electron flow in the vicinity resistance geometry.---}In the following we present numerical results for the components of the magnetic field in Eqs.~(\ref{eq:bx}) and~(\ref{eq:by}), 
evaluated at the position of an array of NV centers, assumed to be aligned at $z= d'$. The magnetic field is  
generated by the 2D current density $\bm{J}(\bm{x})$ in graphene, in the so-called vicinity resistance geometry---see Refs.~\onlinecite{bandurin_science_2016,torre_prb_2015,pellegrino_prb_2016,pellegrino_prb_2017} and below. The graphene sample is modelled as a rectangular stripe of infinite length along the longitudinal direction, $\bm{\hat{x}}$, while it has a finite width 
$W= 2~{\rm \mu m}$ along the transverse direction, $\bm{\hat{y}}$. 
\begin{figure} 
\begin{overpic}[width= \linewidth]{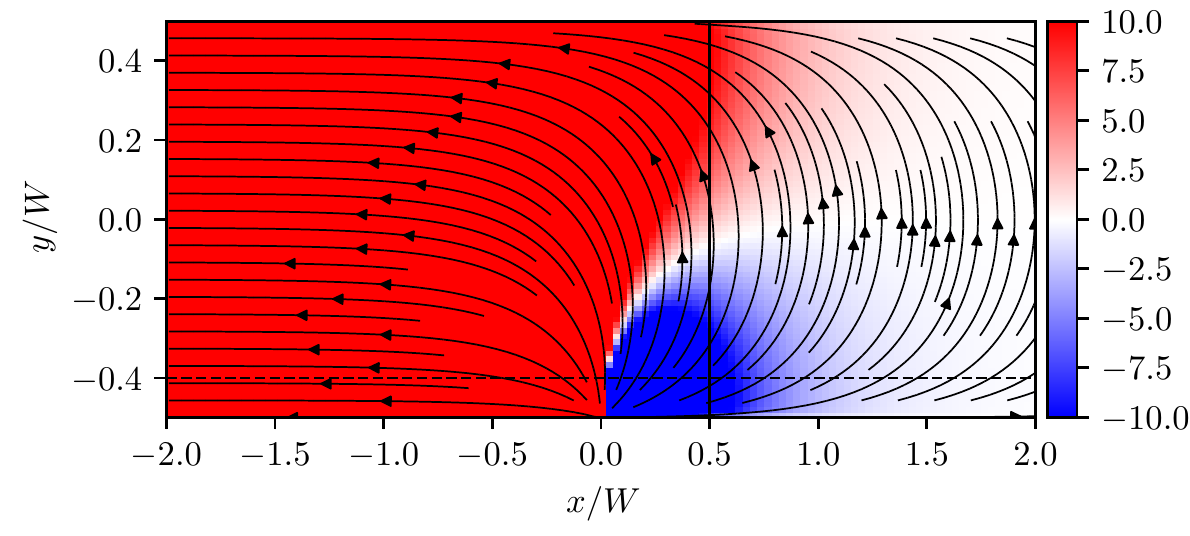}\put(1.85,45){(a)}\end{overpic}
\begin{overpic}[width= \linewidth]{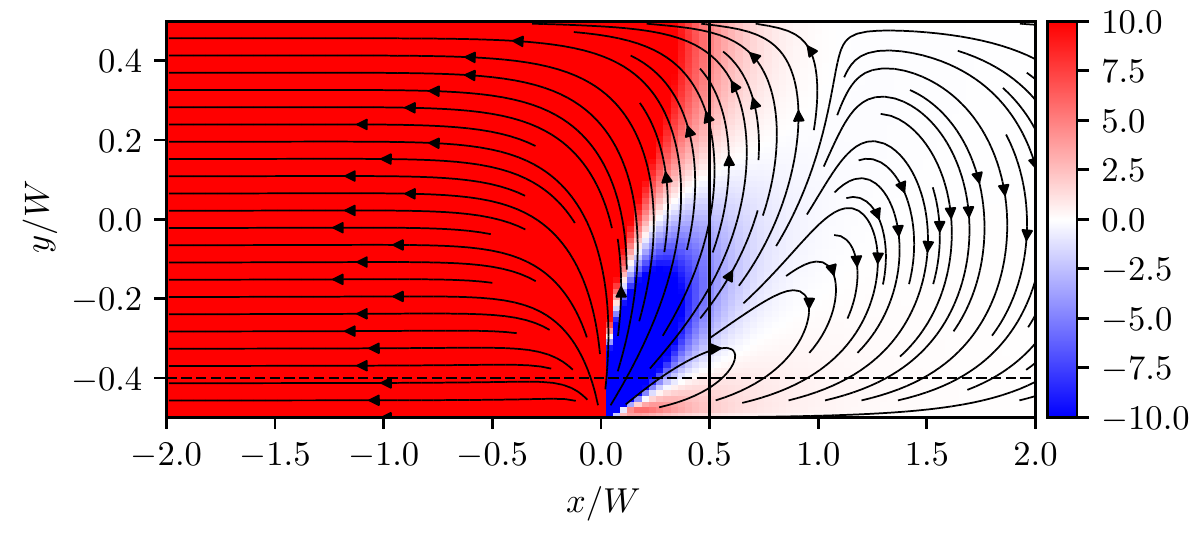}\put(1.85,45){(b)}\end{overpic}
\caption{\label{fig:vicinity-geometry-far-col-by} (Color online) 
Same as in Fig.~\ref{fig:vicinity-geometry-far-col-bx} but for $B_{y}({\bm x}, z=d')$. 
As in Fig.~\ref{fig:vicinity-geometry-far-col-bx}, panel (a) is for the Ohmic case ($D_{\nu}=0$) while panel (b) refers to the viscous case ($D_{\nu}=W/4$).
The vertical and horizontal lines represent the positions where one-dimensional cuts have been taken---see Fig.~\ref{fig:vicinity-geometry-far-col-by-cuts}.} 
\end{figure}
We consider that along the lower edge of the sample (set at $y=-W/2$) there is a point-like current source injecting a current $I$ 
at ${\bm x}_+=(0,y = -W/2)$ and a point-like current drain at ${\bm x}_-=(x_{0},y = -W/2)$, 
while in the remaining points of both edges the normal component of the current density is set at zero, i.e.~$J_y(x,\pm W/2)=0$. 
An additional boundary condition on the tangential component of the current density is required 
at the sample edges~\cite{torre_prb_2015,pellegrino_prb_2016, pellegrino_prb_2017}. Here, we use
the free-surface boundary conditions, i.e.~we impose $[\partial_{y}J_{x}(x,y)+\partial_{x}J_{y}(x,y)]_{y= \pm W/2}= 0$. Below we comment on the impact of a different choice. In this case, and following Ref.~\onlinecite{pellegrino_prb_2016}, we can write
\begin{align} \label{eq:curr-dens-vg}
&\bm{J}(\bm{x}) = I \{ \bm{\nabla} [F(x,y+W/2)- F(x-x_{0},y+W/2)] 
+  \nonumber \\
&  \bm{\nabla} \times \bm{\hat{z}}[G(D_{\nu};x,y+W/2)-G(D_{\nu};x-x_{0},y+W/2)] \}~,
\end{align}
where $F(\bm{x})\equiv \ln[\cosh(\pi x/W)-\cos(\pi y/W)]/(2 \pi)$, and $G(D_{\nu};\bm{x})= 2 D_{\nu}^{2}[\partial_{x} \partial_{y}F(\bm{x})+S(\bm{x})]$, with $S(\bm{x})\equiv \sum_{n= 1}^{\infty}  \sin(n\pi y/W) n\pi \sign(x) e^{-|x|\sqrt{(n \pi/W)^2+1/D_{\nu}^{2}}}/(W^2)$. 

Numerical results have been obtained by setting $d' = 10~{\rm nm}$, and $I = 200~{\rm \mu A}$. 
Here, we compare the magnetic field generated by viscous flow with a realistic value of the viscosity diffusion length, i.e.~$D_\nu=W/4$, with that generated by Ohmic flow, which is mathematically enforced by setting $D_{\nu}= 0$ in Eq.~(\ref{eq:ns}).

Spatial maps of the components $B_x$ and $B_y$ of the magnetic field generated by viscous and Ohmic flows, 
computed from the current density in Eq.~(\ref{eq:curr-dens-vg}) by using Eqs.~(\ref{eq:bx}) and~(\ref{eq:by}),
are reported in Fig.~\ref{fig:vicinity-geometry-far-col-bx} and Fig.~\ref{fig:vicinity-geometry-far-col-by}, respectively. 
By considering the drain at $x_{0} \rightarrow - \infty$, we are able to focus on the electron whirlpool to
the right of the current injector, as seen in Figs.~\ref{fig:vicinity-geometry-far-col-bx}(b) and~\ref{fig:vicinity-geometry-far-col-by}(b).

In Fig.~\ref{fig:vicinity-geometry-far-col-bx} we see that in the viscous case
$B_x({\bm x}, z=d')$ is negative in the regions to the right and the left of the current injector, while the same quantity is positive for $D_{\nu}=0$. A contraction of such regions where $B_x({\bm x}, z=d')$ is negative by increasing temperature (and therefore reducing $D_{\nu}$) signals the occurrence of a smooth crossover from the hydrodynamic to the Ohmic regime. We now note that the effect of a finite viscosity is much more pronounced than what is seen in the color map. Indeed, it is enough to look at 
Fig.~\ref{fig:vicinity-geometry-far-col-bx-cut}, where we present one-dimensional cuts of the 2D spatial map taken 
along the vertical lines $x=W/2$ and $x=3W/2$. We clearly see that, in the Ohmic case, both $B_x$ and $J_y$ are positive definite and concave functions of $y/W$. On the contrary, in the viscous case and in the presence of a whirlpool, the profiles of $B_x$ and $J_y$ have opposite sign and convexity, with respect to the Ohmic case, in an extended range of values of $y/W$, provided that $x$ is sufficiently away from the horizontal position of the current injector ($x=0$)---see Figs.~\ref{fig:vicinity-geometry-far-col-bx-cut}(c) and (d).

\begin{figure}
\begin{overpic}[width= 0.49 \linewidth]{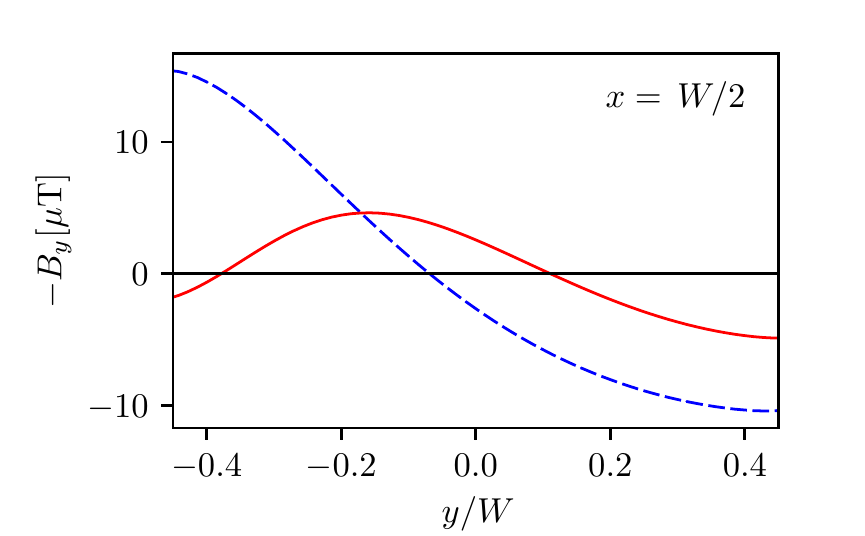}\put(1.6,60){(a)}\end{overpic}
\begin{overpic}[width= 0.49 \linewidth]{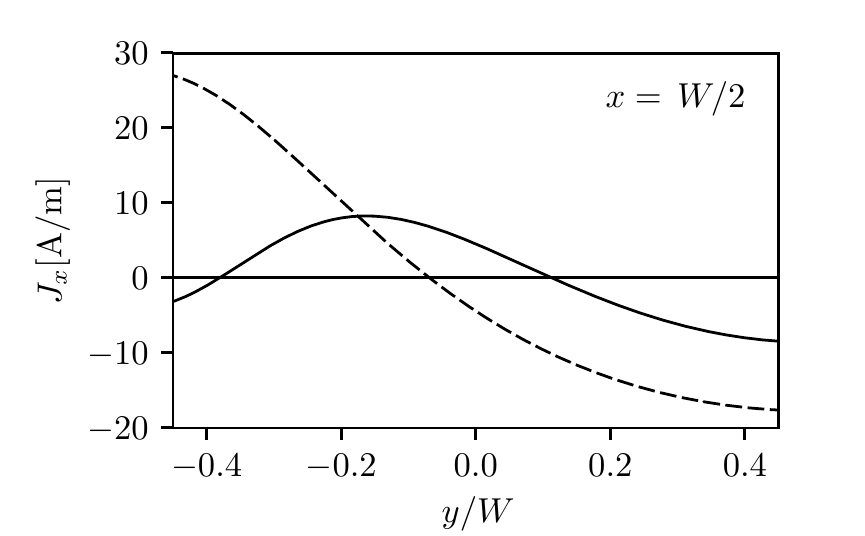}\put(1.8,60){(b)}\end{overpic}
\begin{overpic}[width= 0.49 \linewidth]{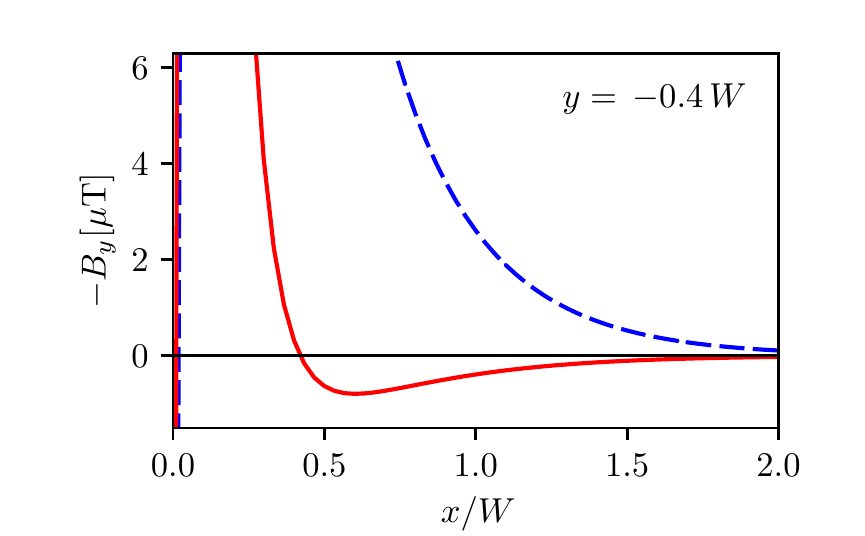}\put(1.6,60){(c)}\end{overpic}
\begin{overpic}[width= 0.49 \linewidth]{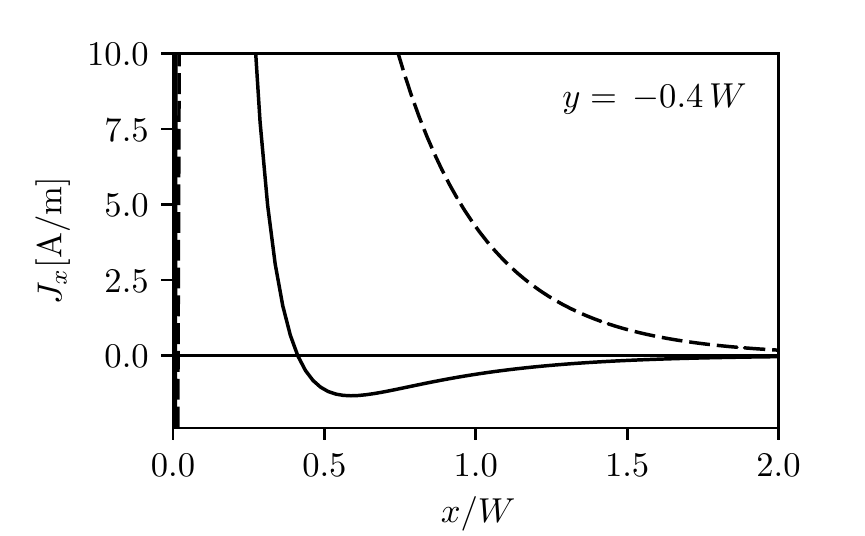}\put(1.8,60){(d)}\end{overpic}
\caption{\label{fig:vicinity-geometry-far-col-by-cuts} 
(Color online) One-dimensional cuts of the 2D spatial maps reported in Fig.~\ref{fig:vicinity-geometry-far-col-by}.
Panel (a) [(b)] illustrates $-B_y({\bm x}, z=d')$ (in units of ${\rm \mu T}$) [$J_x({\bm x})$ (in units of ${\rm A}/{\rm m}$)] evaluated at $x =  W/2$---see solid line in Fig.~\ref{fig:vicinity-geometry-far-col-by}---and as a function of $y/W$.
Panels (c) and (d): same as in panels (a) and (b) but in this case  $-B_y({\bm x}, z=d')$ and $J_x({\bm x})$ are evaluated at $y =  -0.4 \, W$---see dashed line in Fig.~\ref{fig:vicinity-geometry-far-col-by}---and shown as functions of $x/W$. Solid lines in this plot correspond to the viscous case ($D_{\nu}= W/4$), while dashed lines correspond to the Ohmic case ($D_{\nu}=0$).
} 
\end{figure}

A spatial map of $B_y({\bm x}, z=d')$ is reported in Fig.~\ref{fig:vicinity-geometry-far-col-by}, for vanishing---panel (a)---and finite---panel (b)---viscosity diffusion length. 
A clearer signal of viscosity is seen in this figure, in comparison with the map of $B_{x}(\bm{x},z = -d')$
reported in Fig.~\ref{fig:vicinity-geometry-far-col-bx}.
In the viscous case, the negative regions of $B_y({\bm x}, z=d')$ near the current injector are more collimated than in the Ohmic $D_{\nu}=0$ case. Also, positive regions of $B_y({\bm x}, z=d')$ are present to the right of the current injector, where $B_y({\bm x}, z=d')$ is instead negative definite in the Ohmic case. One-dimensional cuts of $B_y({\bm x}, z=d')$ are shown in Fig.~\ref{fig:vicinity-geometry-far-col-by-cuts}, together with $J_x({\bm x})$. Within the shown regions, we clearly see that the profile of $B_y({\bm x}, z=d')$ generated by Ohmic flow is monotonic. 
On the contrary, the presence of a current whirlpool in the viscous case generates a magnetic field with a $B_{y}$ profile featuring clear non-monotonicity. Additionally, backflow in the viscous case generates an additional sign change with respect to the Ohmic case. We conclude that current whirlpools stemming from viscous electron flow in confined geometries determine clear-cut trends in the spatial maps of the generated magnetic field.
\begin{figure}[t]
\begin{overpic}[width= \linewidth]{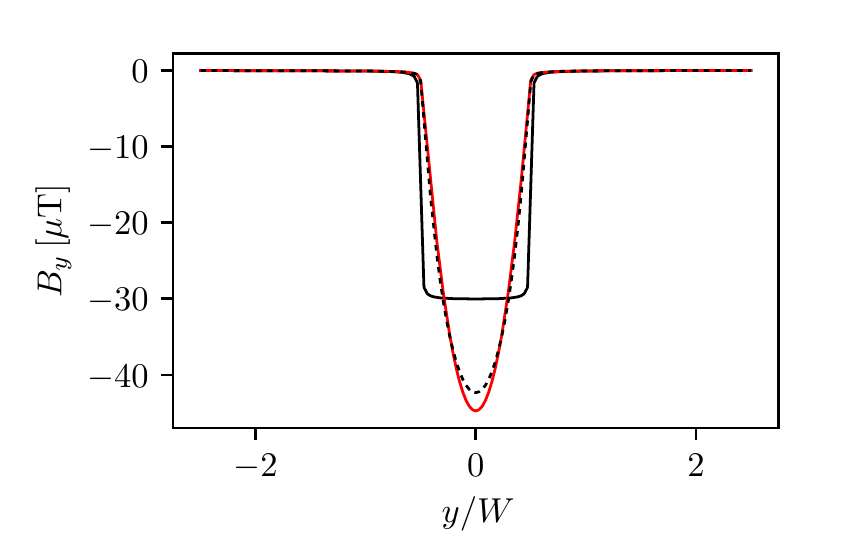}\put(1.85,60){{(a)}}\end{overpic}
\begin{overpic}[width= \linewidth]{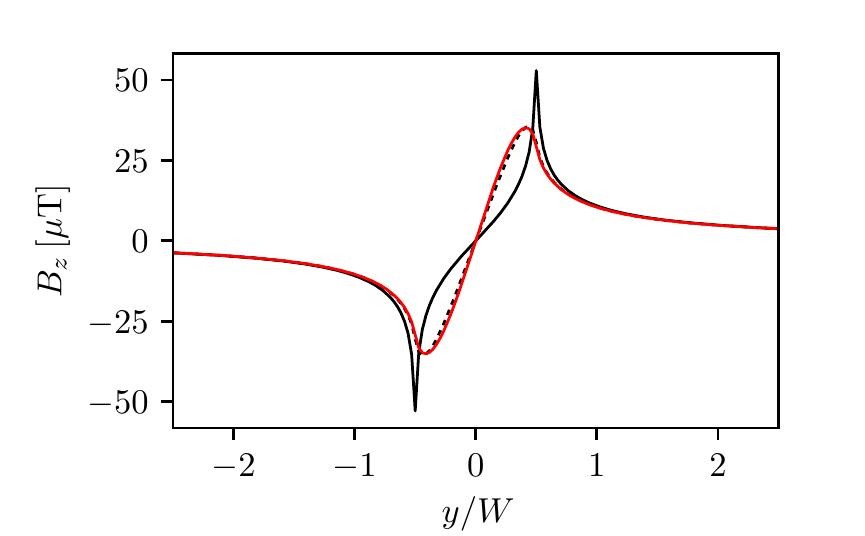}\put(1.85,60){(b)}\end{overpic}
\begin{overpic}[width= \linewidth]{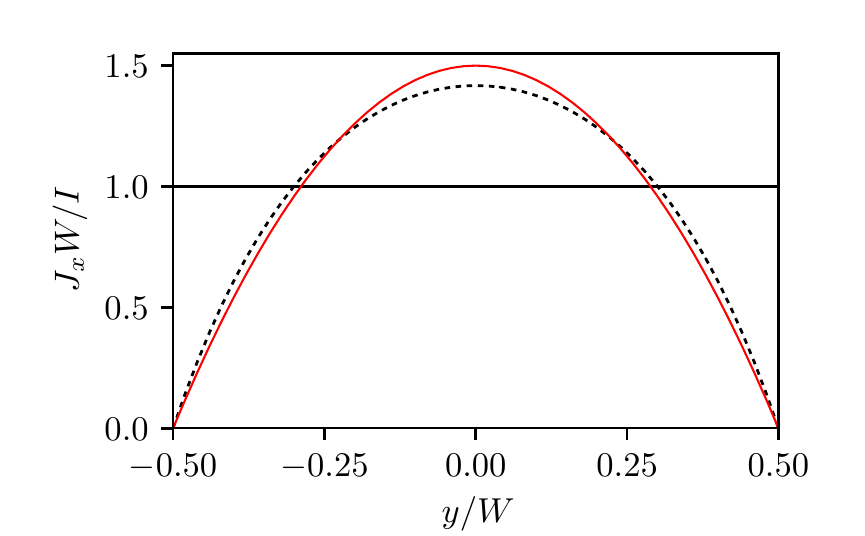}\put(1.85,60){(c)}\end{overpic}
\caption{\label{fig:st-b-and-j} (Color online)
The components $B_y$ and $B_z$ of the magnetic field generated by longitudinal current flow---Eq.~(\ref{eq:curr-dens-tube})---are plotted as functions of $y/W$ in panel (a) and (b), respectively. 
The current density profile (\ref{eq:curr-dens-tube}) is shown in panel (c).
Red solid line: $D_{\nu} = 10 \, W$ and $\ell_{\rm b}=0$; 
black dotted line: $D_{\nu} = W/4$ and $\ell_{\rm b}=0$;
black solid line: $\ell_{\rm b}=\infty$. (With free-surface boundary conditions the result is independent of $D_\nu$.)} 
\end{figure}
\textit{Longitudinal flow.}---We now consider the situation in which no current is injected or extracted laterally at the edges of the graphene sample. In this case, we take current flowing only along the longitudinal direction $\bm{\hat{x}}$.
Following Ref.~\onlinecite{torre_prb_2015}, the current density $\bm{J}(y)= [J_{x}(y),0]$, stemming from the solution of Eqs.~(\ref{eq:continuity}) and~(\ref{eq:ns}), is uniform along $\bm{\hat{x}}$, and reads as following:
\begin{equation} \label{eq:curr-dens-tube}
J_{x}(y)= \dfrac{I}{W} \dfrac{[1-D_{\nu}\cosh(y/D_{\nu}) /\xi]}{[1-2D_{\nu}^{2} \sinh(W/2D_{\nu})/(W \xi)]}~,
\end{equation}
where $\xi \equiv \ell_{\rm b} \sinh(W/2D_{\nu}) + D_{\nu} \cosh(W/2D_{\nu})$. This current density profile has been obtained by describing friction exerted by the edges of the device via the generic boundary conditions~\cite{torre_prb_2015} 
$[\partial_{y}J_{x}(x,y)+\partial_{x}J_{y}(x,y)]_{y= \pm W/2}= \mp J_{x}(x,y= \pm W/2)/\ell_{\rm b}$. Here, the boundary scattering length $\ell_{\rm b}$ allows us to interpolate between the no-slip ($\ell_{\rm b} \rightarrow 0$) and free-surface ($\ell_{\rm b} \rightarrow + \infty$) boundary conditions. Plots of Eq.~(\ref{eq:curr-dens-tube}) for different values of $\ell_{\rm b}$ and $D_{\nu}$ are shown in Fig.~\ref{fig:st-b-and-j}(c). The transition from transverse uniform flow---occurring for free-surface boundary conditions---to Poiseuille flow~\cite{landau_book_fluid}---occurring for no-slip boundary conditions is clearly visible.

The magnetic field generated by the current distribution in Fig.~\ref{fig:st-b-and-j}(c) and evaluated at a distance $d'=10~{\rm nm}$ from the electron fluid---where we assume that the NV centers are placed---is shown in Figs.~\ref{fig:st-b-and-j}(a) and (b). Fig.~\ref{fig:st-b-and-j}(a) shows that
the profile of $B_y$ generated by a transversally-uniform current density $I/W$, obtained by enforcing free-surface boundary conditions, is flat at the center of the graphene sample. In contrast, the profile generated by viscous flow and no-slip boundary conditions displays a parabolic minimum at the center of the graphene sample. The larger the viscosity, the more pronounced the parabolic minimum.
Similarly, Fig.~\ref{fig:st-b-and-j}(b) shows that
the sharpness and the amplitude of the profile of $B_z$ decreases with increasing $D_{\nu}$ by enforcing no-slip boundary conditions, 
while the profile generated by a transversally-uniform current density is sharply peaked at $\pm \overline{y} = \pm \sqrt{W^2+(2 d')^2}/2$.
The detection of the magnetic field generated by a longitudinal flow may therefore enable to determine the suitable boundary conditions for the tangential component of the current density.
Before concluding, we enlighten that in the vicinity resistance geometry, and at odds with the longitudinal geometry, the current pattern is
weakly affected, both quantitatively and qualitatively, by the choice of boundary conditions~\cite{bandurin_science_2016}.

In summary, we have calculated the magnetic field generated by viscous flow in two-dimensional conductors, showing that it displays unambiguous features linked to current whirlpools and backflow near current injectors. We have also shown that the same quantity sheds light on the nature of the boundary conditions describing friction exerted on the electron fluid by the edges of the sample. We believe that our predictions can be tested by carrying out nitrogen-vacancy vector magnetometry on two-dimensional hydrodynamic electron fluids, a technique with spatial resolution that can greatly enrich our understanding of hydrodynamic transport in solid-state systems.
\begin{acknowledgments}
We thank Andrea Tomadin and Jean-Philippe Tetienne for useful discussions. This work has been supported by the European Union's Horizon 2020 research and innovation programme under grant agreement No. 785219 --- GrapheneCore2.
\end{acknowledgments}


\begin{thebibliography}{77}
\bibitem{nonlocal-rel}
The kernel in Eq.~(\ref{eq:kernel-xy}) is peaked around $\bm{x}' = \bm{x}$ and the local relations $B_{x}(\bm{x})^{(0)} \approx (\mu_0/2) J_{y}(\bm{x})$ and $B_{y}(\bm{x})^{(0)} \approx -(\mu_0/2) J_{x}(\bm{x})$ well approximate the exact relation between the components of the magnetic field and the components of the current density. These local relations can be obtained by using a Taylor series expansion of the current density $\bm{J}(\bm{x}')$ around $\bm{x}$, and retaining the lowest order contribution.
\end{thebibliography}
\end{document}